\documentclass[12pt,twoside]{article}
\usepackage{fleqn,espcrc1}
\usepackage{epsfig}

\usepackage[figuresright]{rotating}

\title{
Multiphase transport model for heavy ion collisions at RHIC
\thanks{Supported by NSF Grants PHY-9870038 and PHY-0088934, Welch 
Foundation Grant A-1358, Texas Advanced Research Program Grant 
FY-99-010366-0081, and Arkansas Science and Technology Grant 00-B-14.}}
\author{Zi-wei Lin\address[TAMU]{
Cyclotron Institute and Physics Department, Texas A\&M University, \\
College Station, Texas 77843-3366}, 
Subrata Pal\addressmark[TAMU], C.M. Ko\addressmark[TAMU], 
Bao-An Li\address[ARSU]{
Department of Chemistry and Physics, Arkansas State University, \\
P.O. Box 419, State University, Arkansas 72467-0419}
and Bin Zhang\addressmark[ARSU]}
\begin{document}

\maketitle

\begin{abstract}
Using a multiphase transport model (AMPT) with both partonic and 
hadronic interactions, we study the multiplicity 
and transverse momentum distributions of charged particles such as 
pions, kaons and protons in central Au+Au collisions at RHIC energies. 
Effects due to nuclear shadowing and jet quenching on these 
observables are also studied. We further show preliminary results on 
the production of multistrange baryons from 
the strangeness-exchange reactions during the hadronic stage of
heavy ion collisions.
\end{abstract}

\vspace{1.0cm}

To study heavy ion collisions at the Relativistic Heavy Ion Collider,
in which a quark-gluon plasma is expected to be formed, we have developed
a multiphase transport model that includes both the dynamics of initial
partonic and final hadronic matters \cite{ampt}. In this model, the 
initial space-time information for partons and strings is obtained from 
the HIJING model \cite{hijing}. Scatterings among these minijet partons
are then treated using Zhang's Parton Cascade (ZPC) \cite{zpc}.
After partons stop interacting, they combine with their parent strings 
to form hadrons using the Lund string fragmentation model \cite{lund}
after an average proper formation time of 0.7 fm/c. The dynamics of 
the resulting hadronic matter is then described by a relativistic 
transport model (ART) \cite{art}. 

We first used the model to study heavy ion collisions at SPS energies
and found that it gave smaller numbers of net baryons and kaons 
at midrapidity than those observed in experiments. 
To increase the net baryon at midrapidity, we introduced both the 
popcorn mechanism of baryon-antibaryon production in the Lund
string fragmentation and baryon-antibaryon production from and 
annihilation to mesons in the ART model. The kaon number was increased 
by adding in the ART model the production and destruction
of $K^*$ resonances, and by adjusting the two parameters 
in the splitting function used in the Lund string fragmentation.

\begin{figure}[htb]
\begin{minipage}[t]{77mm}
\centerline{\includegraphics[width=3.in,height=3.2in,angle=0]{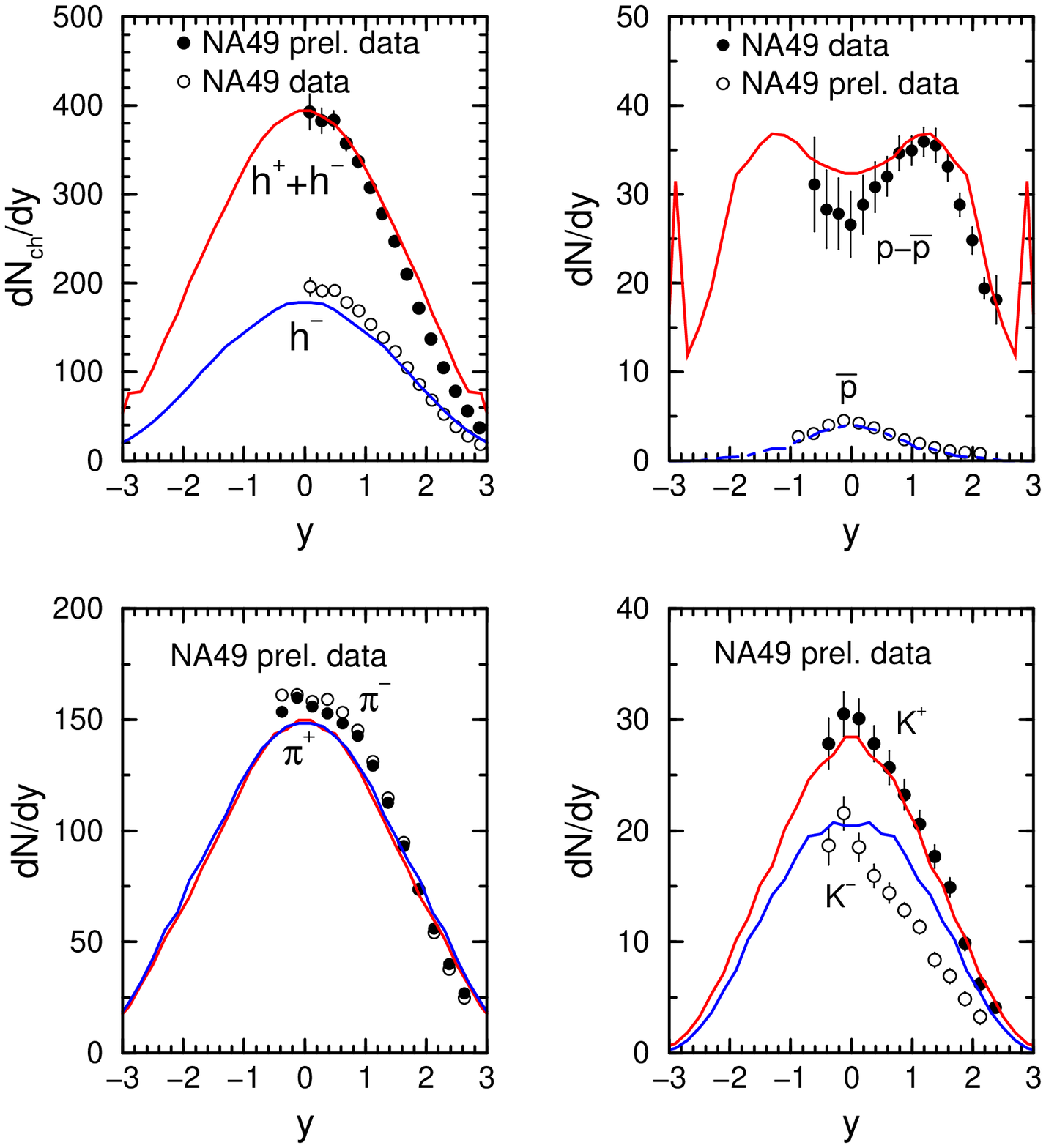}}
\vspace{-0.6cm}
\caption{Rapidity distributions at SPS.} 
\label{fig1}
\end{minipage}
\hspace{\fill}
\begin{minipage}[t]{77mm}
\centerline{\includegraphics[width=3.in,height=3.22in,angle=0]{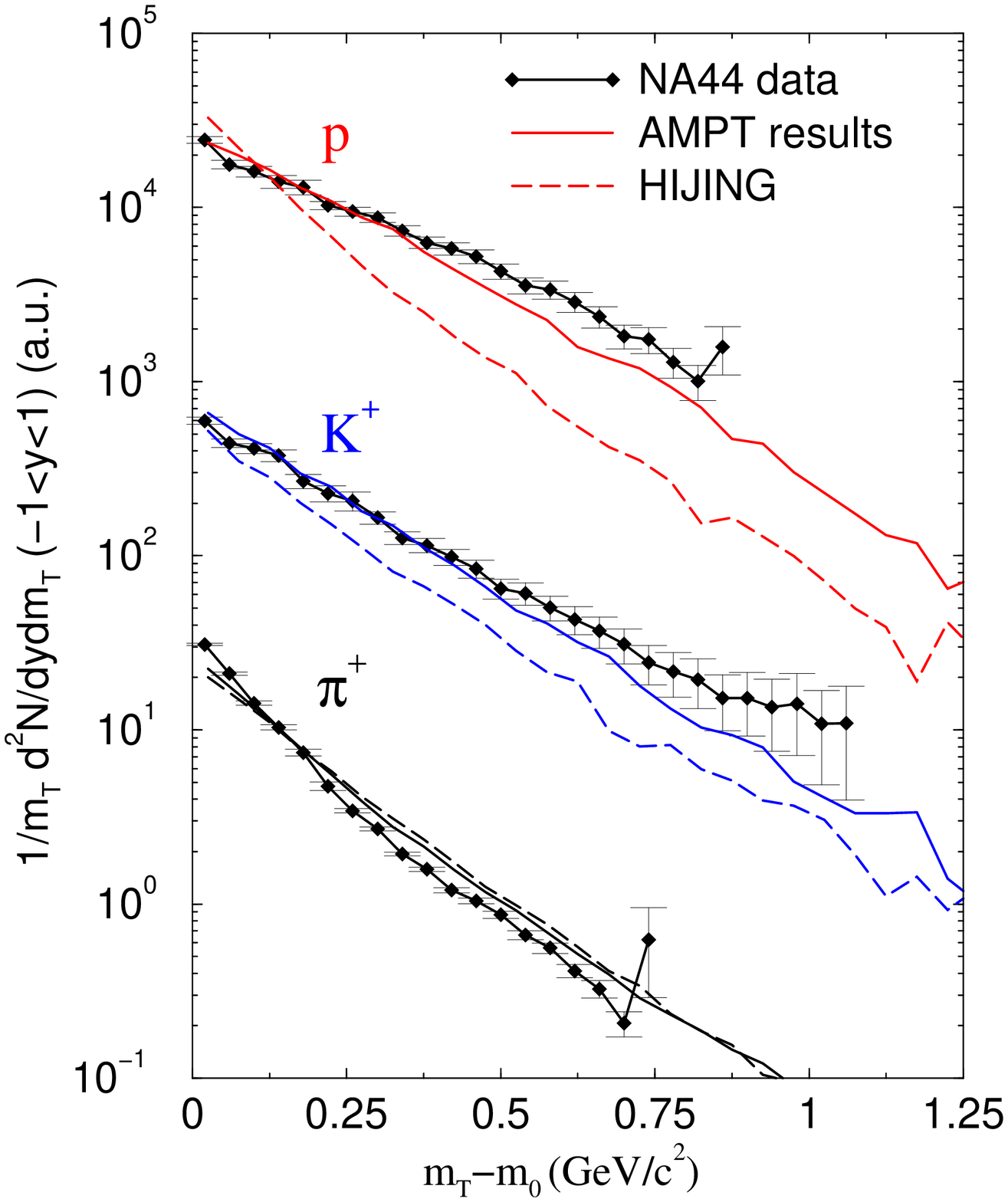}}
\vspace{-0.6cm}
\caption{Transverse momentum spectra at SPS.} 
\label{fig2}
\end{minipage}
\end{figure}

Results from the improved AMPT model for central Pb+Pb collisions at
$158A$ GeV from the SPS, corresponding to impact parameters of $b\leq 3$ fm,
are shown in Fig. \ref{fig1} for the
rapidity distributions and in Fig. \ref{fig2} for the 
transverse momentum spectra of charged particles including
pions, kaons, protons, and antiprotons. We see that the theoretical
results shown by solid curves agree reasonably with the experimental 
data \cite{data}. On the other hand, the HIJING model with default 
parameters, shown by dashed curves in Fig. \ref{fig2},
underpredicts the inverse slopes of the transverse momentum 
spectra for kaons and protons in these collisions. 
Final state hadronic scatterings are thus important in describing
the transverse momentum spectra.

\begin{figure}[htb]
\vspace{-0.6cm}
\begin{minipage}[t]{77mm}
\centerline{\includegraphics[width=3.in,height=3.5in,angle=0]{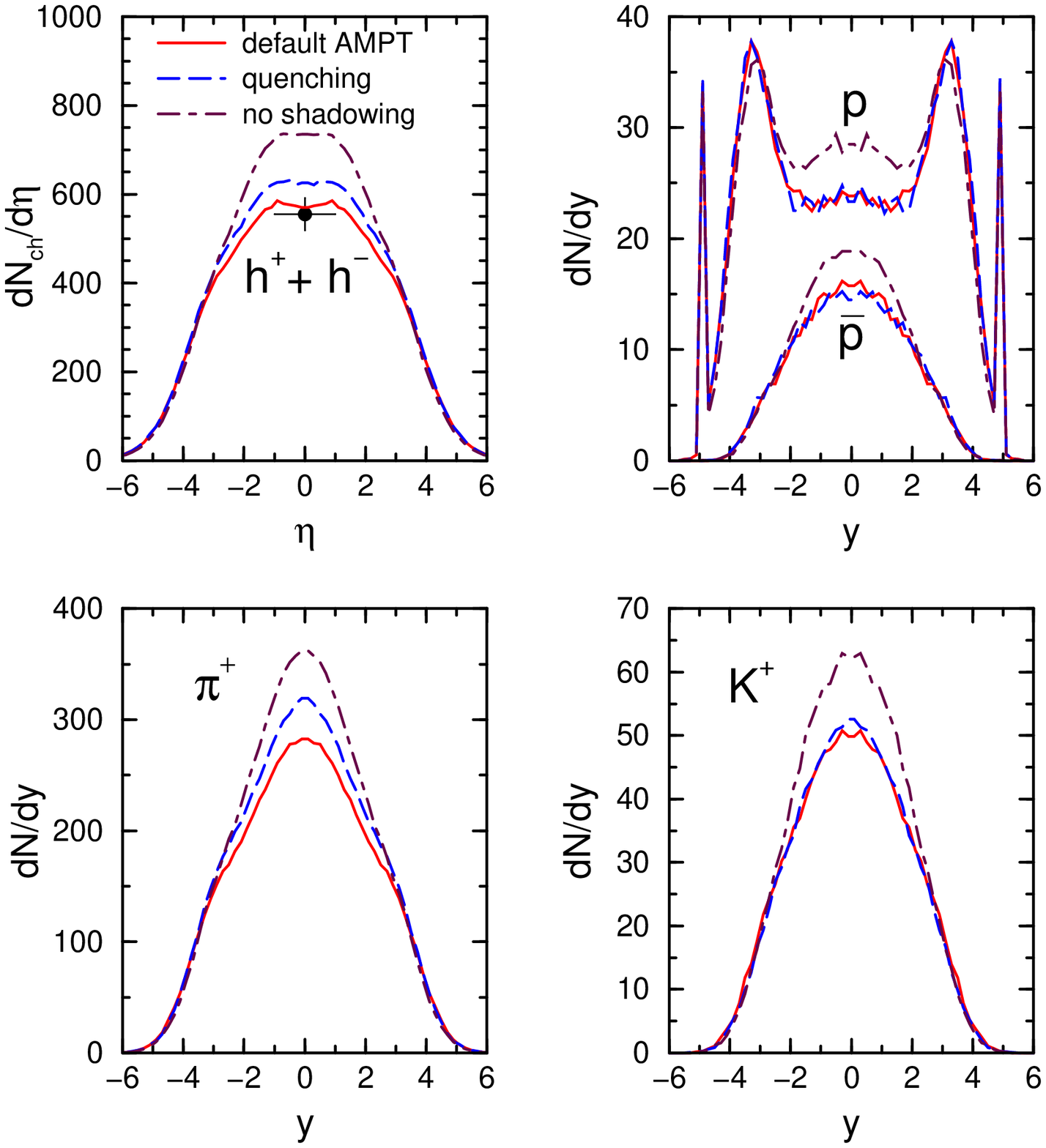}}
\vspace{-0.6cm}
\caption{Rapidity distributions at RHIC.} 
\label{fig3}
\end{minipage}
\hspace{\fill}
\vspace{-0.6cm}
\begin{minipage}[t]{77mm}
\centerline{\includegraphics[width=3.in,height=3.15in,angle=0]{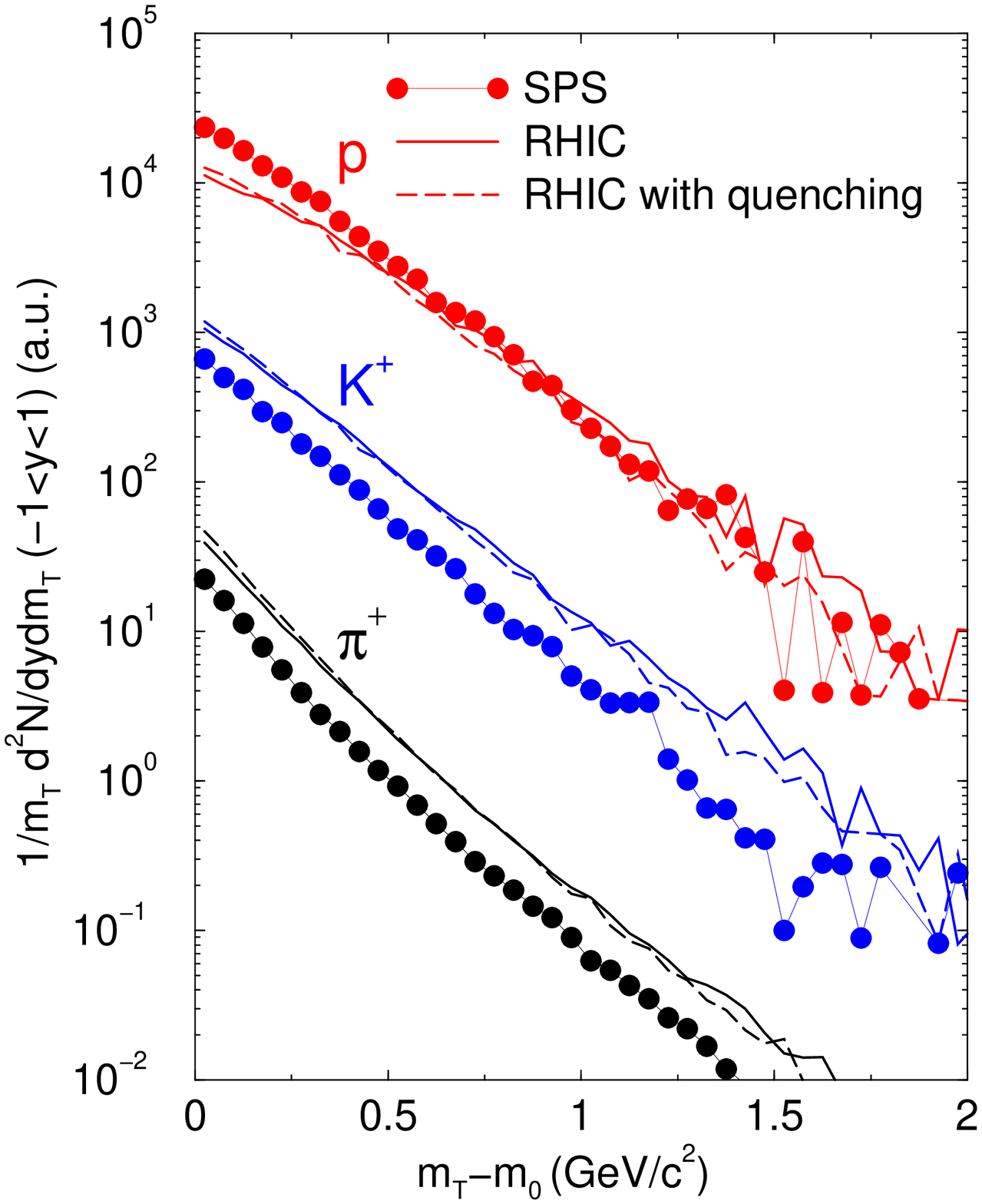}}
\vspace{-0.6cm}
\caption{Transverse momentum spectra at RHIC.}
\label{fig4}
\end{minipage}
\end{figure}

With parameters in the AMPT model constrained by experimental data from 
heavy ion collisions at 
SPS energies, we then studied heavy ion collisions at RHIC energies. We first 
show in Fig.~\ref{fig3} by the solid curve the pseudorapidity distribution 
of total charged particles in central ($b\leq 3$ fm) Au+Au collisions at 
$130A$ GeV. The theoretical result is consistent with the data from the 
PHOBOS collaboration \cite{Phobos00} shown in the figure by the full circle.
We note that the total charged particle multiplicity increases 
appreciably without final hadronic scatterings but is hardly affected 
by partonic scatterings. The latter is partly due to the absence of
inelastic scattering in the ZPC model. To model the effect of parton
energy loss due to inelastic scatterings, we introduce the jet quenching 
as in default HIJING. In Fig. \ref{fig3}, we show by dashed curves the results 
from AMPT model with a jet quenching of $dE/dx=1$ GeV/fm. 
Also shown in Fig. \ref{fig3} are the results obtained by neglecting
nuclear shadowing on parton production in the AMPT model. 
In both cases, the total charged particle multiplicity is 
larger than that from the default AMPT model. The PHOBOS data is thus 
consistent with a significant nuclear shadowing effect but a rather 
weak jet quenching.  Our predictions for the multiplicities of pions, kaons, 
protons and antiprotons are also shown by solid curves in Fig.~\ref{fig3}. 
It is seen that the $\bar p/p$ ratio is significantly increased in comparison 
with that in central Pb+Pb collisions at SPS. 
 
In Fig.~\ref{fig4}, we show by solid curves the transverse momentum spectra 
of pions, kaons, and protons in central Au+Au collisions at $130A$ GeV.   
For comparison, we also show by circled curves the default 
AMPT results for central Pb+Pb collisions at SPS as already shown in 
Fig.~\ref{fig2} and by dashed curves the AMPT results with a jet quenching 
of $dE/dx=1$ GeV/fm. We find that all three particles, especially protons, 
have larger transverse momenta at RHIC than at SPS.  
As expected, jet quenching suppresses moderately the yields of these 
particles at high transverse momenta. 

We have also extended the AMPT model to include strangeness-exchange reactions 
in the hadronic matter in order to study the production of multistrange 
baryons in relativistic heavy ion collisions.
This is important for understanding whether one can use
enhanced production of multistrange particles as a signal for the 
formation of quark-gluon plasma in relativistic heavy ion collisions 
\cite{rafelski}. The cross sections for $\Lambda$ and $\Sigma$ 
productions from $K^-$ and nucleon interactions are known empirically. 
From the parameterized cross sections for $K^- p \rightarrow \pi^0 \Lambda$ 
and $\pi^0 \Sigma^0$, and $K^- n\rightarrow \pi^- \Lambda^0$ \cite{Cugn90}, 
the isospin-averaged cross section of hyperon production from 
$\bar K N \rightarrow \pi Y$ can be obtained, and it is $3/2$ times 
their sum. The cross sections for $\Xi$ and $\Omega$ production, i.e.,  
$\bar K \Lambda \rightarrow \pi \Xi$ and $\bar K \Xi \rightarrow \pi \Omega$,
are unknown. In our preliminary study, we assume that they are the same
as that for $\bar K N \rightarrow \pi Y$. 

\begin{table}[htb]
\caption{
Strange particle yields at midrapidity from AMPT 
at SPS and RHIC ($\sqrt s=130A$ GeV).}
\begin{tabular}{llllllll}
\hline
& $K^+$ & $K^-$ & $\Lambda$ & $\Xi^-$ & $\Omega^-$ 
& $\Xi^-/\Lambda$ & $\Omega^-/\Xi^-$ \\
\hline
SPS  & 28 & 19 & 12 & 1.1 & 0.19 & 0.093 & 0.16 \\
RHIC & 50 & 45 & 13 & 2.0 & 0.43 & 0.16  & 0.22 \\
RHIC ($P_s=P_u$) & 68 & 61 & 13 & 5.1 & 1.9 & 0.38 & 0.37 \\
\hline
\end{tabular}\\[2pt]
\end{table}

In Table 1, we show the abundance of strange hadrons at midrapidity 
($|y|<0.5$) from the AMPT model that includes the above 
strangeness-exchange reactions and their inverse reactions.
We see that our results for central Pb+Pb collisions at SPS 
($\sqrt s=17A$ GeV) are consistent (within 50\%) with the large 
enhancement of multistrange particle production observed in the experiments. 
We note that a similar conclusion was reached in Ref. \cite{Vance01} 
using a different assumption on the cross sections of strangeness-exchange
reactions. 
Although the AMPT model includes a partonic stage, 
no strange quark production is considered at present. To simulate the
effect of strange particle production during the partonic stage of heavy ion 
collisions, we let strange quarks to have the same production probability
as light quarks in the Lund string fragmentation. The results from
such a scenario are shown in the last row of Table 1. 
It is seen that this leads to a significant enhancement of multistrange 
particle production, although its effect on particles with single strangeness,
such as $K$ and $\Lambda$, is much smaller. Our results thus imply that 
enhanced production of multistrange particles in relativistic heavy 
ion collisions remains a possible signal for the initial quark-gluon plasma 
in spite of multistrange particle production from the hadronic matter.

To summarize, a multiphase transport model has been developed for
heavy ion collisions at RHIC energies. Comparisons of the theoretical
results with the PHOBOS data on total charged multiplicity indicate
that there is a significant nuclear shadowing but a weak jet quenching 
in the initial stage of collisions. The predicted transverse momentum 
spectra of particles are found to have larger inverse slopes at RHIC 
than at SPS. We have also studied the effect of strangeness-exchange 
reactions on the production of multistrange particles and found that  
the observed enhancement in heavy ion collisions at SPS can be largely 
explained. For heavy ion collisions at RHIC, strangeness-exchange 
reactions also enhance the yield of multistrange particles. Allowing 
enhanced production of strange particles during hadronization
further increases significantly the ratios of multistrange to 
single-strange particles. Multistrange particle production is 
thus a sensitive probe to the early dynamics of heavy ion collisions.


\begin{thebibliography}{9}
\bibitem{ampt} 
B. Zhang, C.M. Ko, B.A. Li and Z.W. Lin, Phys. Rev. C 61 (2000) 067901;
Z.W. Lin, S. Pal, C.M. Ko, B.A. Li and B. Zhang, eprint No. nucl-th/0011059.
\bibitem{hijing} 
X.N. Wang and M. Gyulassy, Phys. Rev. D 44 (1991) 3501.  
\bibitem{zpc} 
B. Zhang, Comp. Phys. Comm. 109 (1998) 193.
\bibitem{lund} 
T.  Sj\"{o}strand, Comp. Phys. Comm. 82 (1994) 74. 
\bibitem{art} 
B.A. Li and C.M. Ko, Phys. Rev. C 52 (1995) 2037. 
\bibitem{data}
I.G. Bearden {\em et al.} (NA44 Collaboration), 
Phys. Rev. Lett. 78 (1997) 2080;
H. Appelsh\"auser {\em et al.} (NA49 Collaboration), 
Phys. Rev. Lett. 82 (1999) 2471;
F. Sikl\'er {\em et al.} (NA49 Collaboration), 
Nucl. Phys. A 661 (1999) 45c; 
S. Jeon and J. Kapusta, Phys. Rev. C 63 (2001) 011901.
\bibitem{Phobos00}
B.B. Back {\em et al.}, PHOBOS Collaboration, Phys. Rev. Lett. 85 (2000) 3100. 
\bibitem{rafelski}
J. Rafelski and B. M\"uller, 
Phys. Rev. Lett. 48 (1982) 1066; 56 (1986) 2334 (E).
\bibitem{Cugn90} 
J. Cugnon, P. Deneye and J. Vandermeulen, Phys. Rev. C 41 (1990) 1701. 
\bibitem{Vance01} 
S. Vance, J. Phys. G 27 (2001) 603. 
\end{thebibliography}
\end{document}